\documentclass[superscriptaddress,aps,prx,twocolumn,showpacs,nofootinbib,longbibliography,notitlepage,floatfix]{revtex4-2}
\usepackage{etex}
\usepackage{amsmath,amssymb,amsthm}
\usepackage[colorlinks=true,citecolor=blue,urlcolor=blue]{hyperref}
\usepackage[pdftex]{graphicx}
\usepackage{times,txfonts}
\usepackage{braket}
\usepackage{color}
\usepackage{natbib}
\usepackage{amsmath,blkarray}
\usepackage{mathtools}
\usepackage{ulem}
\usepackage{latexsym}
\usepackage{tabularx, booktabs}
\usepackage{graphics,epstopdf}
\usepackage{graphicx}
\usepackage{float}
\usepackage{amsfonts}
\usepackage{tikz}
\usetikzlibrary{quantikz}
\usepackage{color,soul}

\newcommand{\be}{\begin{equation}}
\newcommand{\ee}{\end{equation}}
\newcommand{\ba}{\begin{eqnarray}}
\newcommand{\ea}{\end{eqnarray}}

\usepackage{multirow}
\usepackage{appendix}
\usepackage{url}

\begin{document}

\title{Experimental Demonstration of Quantum Tunneling in IBM Quantum Computer}

\author{Narendra N. Hegade}
\thanks{These authors contributed equally}
\affiliation{Department of Physics, National Institute of Technology Silchar, Silchar, 788010, India }

\author{Nachiket L. Kortikar}
\thanks{These authors contributed equally}
\affiliation{Department of Physics, Freie Universit$\ddot{a}$t, Berlin, Physikinstitut der FU, Arnimallee 14, 14195 Berlin, Germany }

\author{Bikramaditya Das}
\affiliation{Department of Chemistry, University of Calcutta, 92 APC Road, 700009, Kolkata, West Bengal, India}

\author{Bikash K. Behera}
\author{Prasanta K. Panigrahi}
\email{pprasanta@iiserkol.ac.in}
\affiliation{Department of Physical Sciences, Indian Institute of Science Education and Research Kolkata, Mohanpur, 741246, West Bengal, India}


\begin{abstract}
Quantum computers are the promising candidates for simulation of large quantum systems, which is a daunting task to perform in a classical computer. Here, we report the experimental realization of quantum tunneling of a single particle through different types of potential barriers by performing digital quantum simulations using IBM quantum computers. We consider two and three-qubit systems to visualize the tunneling process and illustrate its unique quantum nature. We observe the tunneling and oscillations of the particles in a step-well, double-well, and multi-well potentials through our experimental results. One may extend the proposed quantum circuits and simulation techniques used here for observing the tunneling phenomena for multi-particle systems in different potentials.
\end{abstract}

    
\maketitle

\section{introduction}
Quantum simulation is one of the problems that a quantum computer could perform more efficiently than a classical computer as it provides a significant improvement in computational resources \cite{lloyd1996universal,abrams1997simulation,zalka1998simulating,berry2007efficient,deutsch1992rapid,simon1997power,gerjuoy2005shor}. It has been applied to a wide range of areas of physics like quantum many-body theory  \cite{tseng1999quantum, negrevergne2005liquid, peng2009quantum, feng2013experimental}, quantum entanglement \cite{zhang2011experimental, peng2005quantum}, quantum phase transitions, molecular physics \cite{aspuru2005simulated, lanyon2010towards, du2010nmr} etc. Algorithms have been used in simulating many quantum field theoretic problems \cite{buchler2005atomic, zohar2011confinement, cirac2010cold, kapit2011optical, bermudez2010wilson, lepori20103+, maeda2009simulating, casanova2011quantum, douccot2004discrete, johanning2009quantum, bender2018digital}, where Hamiltonian of the system splits into kinetic and potential energy operators which are then simulated using Trotter's formula \cite{jordan2012quantum, byrnes2006simulating}. Experimental realizations of quantum simulation have already been made in systems like NMR \cite{tseng1999quantum, negrevergne2005liquid, peng2009quantum, peng2005quantum, du2010nmr, somaroo1999quantum, khitrin2001nmr, brown2006limitations}, ion-trap \cite{friedenauer2008simulating, gerritsma2010quantum, gerritsma2011quantum}, atomic \cite{edwards2010quantum} and photonic \cite{aspuru2012photonic} quantum computers.

Quantum tunneling \cite{balantekin1998quantum} acts as one of the exciting and unique fundamental phenomena in quantum mechanics. It has been observed in superconducting Cooper pairs \cite{leggett1980diatomic} and in modern technologies such as narrow p-n junctions \cite{esaki1960new} and scanning tunneling microscope \cite{tersoff1985theory}. Some important puzzling phenomenon in science such as lattice quantum chromodynamics \cite{kogut1983lattice} can be solved using this tunneling simulation approach \cite{jensen1991methodology, shen2008variational, jensen1993numerical}. A number of digital simulation on quantum tunneling \cite{feng2013experimental} has been performed on classical computers and photonic systems \cite{lu2016topological}. This type of simulation has remained untested in a quantum computer due to the requirement of large number of ancillary qubits and quantum gates. Recently, an algorithm proposed by Sornborger \cite{cirstoiu2020variational} illustrates the simulation with no ancillary qubits and a small number of quantum gates, which motivates the possibility of simulating in today's quantum computer consisting of a few number of qubits. It is already demonstrated the tunneling effect for two-qubit and three-qubit systems using NMR quantum computer \cite{jones1998implementation, peng2005quantum}. Ostrowski \cite{ming2015integrated} has also explicated this process on a rectangular potential by digital simulation. 

In this work, we illustrate the simulation of quantum tunneling phenomenon using IBM’s 5-qubit quantum computer “ibmqx4”, "ibmqx2" and 14-qubit quantum computer “IBM Q 14 Melbourne”. Using two and three qubit systems by using a set of CNOT, Hadamard and controlled phase gates, we were able to simulate the tunneling process of a single particle in a step potential, double-well potential and
multi-well potentials. We have utilized the IBM quantum experience’s Qiskit package to simulate the tunneling Hamiltonian, using which a number of research works have been performed. A lot of recent works in quantum computation, quantum simulation and quantum information pursue the use of IBM quantum experience \cite{qsim1,qsim2,qsim3,qsim4,qsim5,qsim6,qsim7,qsim8,qsim9,qsim10,qsim11,qsim12,qsim13,qsim14,qsim15}.

The manuscript is arranged as follows. In the following section Sec.~\ref{Sec-II}, we present the theoretical protocol explaining the three main steps on a quantum simulation that is initial state preparation, time evolution, and measurement. We then show how to decompose the time evolution of a Hamiltonian into quantum circuits using set of single qubit and entangling gates. We also give analytical solutions for the evolution of particle in different potential wells.  In Sec.~\ref{Sec-III} we present the circuit construction for two and three-qubit simulations implementing the decomposed operators from Sec.~\ref{Sec-II} using Hadamard and Controlled-phase gates. Sec.~\ref{Sec-IV} consists of results from digital simulations on ibmqx2, ibmxq4 and IBM Q 14 Melbourne in free space, step-well, and double square-well potentials implemented by constructing circuits using two-qubits and multi-well potential implemented by constructing circuits using three-qubits. We also present the results of tomography and fidelity as a function of time steps for different devices namely IBM Q 14 Melbourne, ibmqx2, and ibmqx4. In Sec.~\ref{Sec-V} we discuss our results and conclude.

\section{Theoretical Protocol \label{Sec-II}}
\begin{figure}
\centering
\includegraphics[width=\linewidth]{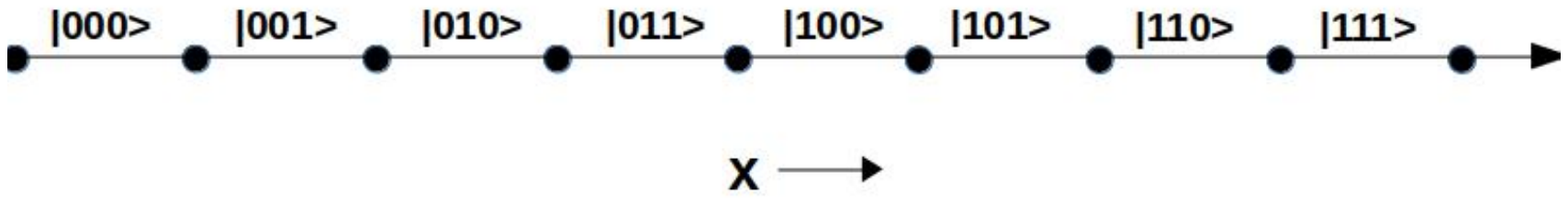}
\caption{Binary representation of a wavefunction on a 1D lattice for 3-qubit case. The 3-qubit quantum states \(\ket{000}, \ket{001},..., \ket{111}\) represent the binary form of the states \(\ket{0}, \ket{1},..., \ket{7}\)
respectively.}
\label{Fig1.pdf}
\end{figure}

A digital quantum simulator \cite{las2014digital} is a controllable quantum system \cite{schirmer2001complete} that can efficiently simulate the dynamics of any other quantum system with local interactions. The digital quantum simulation \cite{schoelkopf2008wiring} consists of three main steps: initial state preparation, time evolution, and measurement. In this work, we are particularly interested in the time evolution of the system, so our main idea is to decompose the Hamiltonian of the system in terms of single and two-qubit gates. The Schrodinger’s equation, for a single particle moving in an one-dimensional space, is expressed as

\begin{eqnarray}
i\frac{\partial}{\partial t} \ket{\psi(x,t)} = \hat{H} \psi(x,t).
\end{eqnarray}
Here, $\hat{H}$ = $(\hat{K} + \hat{V})$ , $\hat{K}$ and $\hat{V}$ are kinetic and potential energy operators, respectively. Here, we
set the value of $\hbar$ to $1$ throughout the manuscript. The time evolution of the wave function of the system can be given as
\begin{eqnarray}
\ket{\psi(x,t+\Delta t)} = e^{-i \hat{H} \Delta t}\ket{\psi(x,t)} = e^{i(\hat{K}+\hat{V})\Delta t}\ket{\psi(x,t)}.\nonumber
\end{eqnarray}
Using first-order Suzuki-Trotter formula \cite{de1983applications, hatano2005finding, sornborger1999higher}, the sum of exponential operator can be decomposed as follows
\[e^{-i \hat{H} \Delta t} \approx e^{-i \hat{K}\Delta t}e^{-i\hat{V}\Delta t}.\]
This approximation introduces an error of the order $O(\Delta t^2)$. For better approximation, one can consider higher order trotter formula. 
We discretize the continuous co-ordinate space $x$ on a lattice (with spacing \(\Delta l\)) within the boundary region (\(0 \leq x \leq L\)) with a periodic boundary condition $\ket{\psi(x+L,t)} = \ket{\psi(x,t)}$. The wave
function then can be mapped to a n-qubit register, 
\begin{eqnarray}
\ket{\psi(x,t)} = \sum_{k=0}^{2^n -1}\psi(x_k,t)\ket{k}.
\end{eqnarray}
Here \(\ket{k}\) represents the particle location corresponding to binary number k, and \(x_k = (k +\frac{1}{2}\Delta l)\), where \(\Delta l = \frac{L}{2^n}\). The mapping given here can be a good approximation for large value
n. As an example, for n=3 qubits, the binary representation of a wavefunction on a 1D lattice is
represented in Figure \ref{Fig1.pdf}. Here, the wave function mapped to the 3-qubit register can be written as
\begin{eqnarray}
\ket{\psi(x,t)} = [\psi(x_0,t)\ket{0} +\psi(x_1,t)\ket{1}\nonumber\\
+ \psi(x_2,t)\ket{2}+...+ \psi(x_7,t)\ket{7} ]\;,
\end{eqnarray}
where \(\sum_{k=0}^7(\psi(x_k,t)^2 =1\)  and  \(x_k = \frac{L}{8}\). The quantum states \(\ket{0},\ket{1},...,\ket{7}\) are mapped into a 3-qubit quantum states as \(\ket{000},\ket{001},...,\ket{111},\) respectively.

\subsection{Implementing Potential Energy Operator}
The potential energy operator plays a key role for introducing and studying various types of potential structures. Here, we implement step-well, double-well and multi-well potentials with the help of single qubit rotation operators without ancillary qubits. For step potential, we apply a single-qubit Z-rotation gate on the highest order qubit,
\[e^{-i\hat{V}\Delta t} = e^{-iv\sigma_z^{n-1}\Delta t} = e^{-iv\sigma_z\Delta t}\otimes I \otimes I...\otimes I\;,\]
where $v$ corresponds to the magnitude of the potential barrier, \(\sigma_z\) is the Pauli-Z matrix, and \((n-j)\) indicates the application of the operator
\((e^{-i v \sigma_z \Delta t})\) on the $j$th qubit. The double-well potential can be implemented by applying the Z-rotation gate on the second highest order qubit,\\
\[e^{-i \hat{V}\Delta t} = e^{-iv\sigma_z^{n-2}\Delta t} = I \otimes e^{-iv\sigma_z\Delta t}\otimes I...\otimes I.\]\\
Similarly, by applying Z-rotation operator on the next highest order qubits, we can implement multi-well potentials,
\[e^{-i \hat{V}\Delta t} = e^{-iv\sigma_z^{n-3}\Delta t} = I \otimes I \otimes e^{-iv\sigma_z\Delta t}\otimes I...\otimes I.\]
It can be observed that just one single qubit operation can reduce the complexity in the quantum circuit by replacing a large number of gates and ancillary qubits \cite{gerritsma2011quantum}.

\subsection{Implementing Kinetic Energy Operator}
Here, the kinetic energy operator $\hat{K}$ can be expressed in terms of momentum operator as, \(\hat{K} = \frac{\hat{p}^2}{2m}\). For finding the quantum circuit for kinetic energy operator, we discretize the wave function of momentum as \cite{zhang2009direct, zhang2011experimental}
\begin{eqnarray}
\ket{\chi(p,t)}= \sum_{l=0}^{2^n -1}\chi(p_l,t)\ket{l},
\end{eqnarray}
where \(\chi(p, t)\) is the wave function in the momentum basis. The eigenvalues of momentum operator is given by \cite{farout2020exact}
\begin{equation}\label{eqnx}
p_{l}=
\begin{cases}
\frac{2\pi}{2^n}l & 0\leqslant l\leqslant2^{n-1}
\\
\frac{2\pi}{2^n}(2^{n-1}-l) & 2^{n-1} < l < 2^{n}.
\end{cases}
\end{equation}
In the momentum representation, the diagonal operator $\hat{P}$ is written as
\begin{eqnarray}\label{phat}
 \hat{P} = \sum_{l=0}^{2^{n - 1}} \frac{2\pi l}{2^n}\ket{l}\bra{l} + \sum_{2^{n-1}+1}^{2^n - 1} \frac{2\pi (2^{n-1} - l)}{2^n}\ket{l}\bra{l}.
\end{eqnarray}
The kinetic energy operator is diagonal in the momentum representation. It can be written in the co-ordinate representation by using quantum fourier transform as follows \cite{lloyd1996universal, zalka1998simulating}
\begin{eqnarray}\label{eqnm}
e^{-i(\hat{K}+\hat{V})\Delta t}\approx & (QFT) e^{-i\frac{\hat{p}^2}{2m}\Delta t} (QFT^{-1}) e^{-i\hat{V}\Delta t} \nonumber \\
\approx & (QFT)\hat{D}(QFT^{-1})\hat{P}\;,
\end{eqnarray}
where 
\begin{eqnarray}QFT =\frac{1}{\sqrt{2^n}}\sum_{{l,k}=0}^{2^n - 1} \frac{2\pi ilk}{2^n}\ket{l}\bra{k}\;,\\
D=e^{-i\frac{{\hat{p}}^2}{2m}}\;, \quad P=e^{-i \hat{V}\Delta t}.
\end{eqnarray}
The equivalent quantum circuit for the Fourier transformation operator $QFT$ \cite{zhang2013watermark} can be realized
using a series of Hadamard and controlled-phase gates. Thus, after a small time interval \(\Delta t\), the
time evolution of the system can be implemented as
\begin{equation}
\sum_{k=0}^{2^n  - 1}\ket{\psi(x_k,t+\Delta t)}
= (QFT)D(QFT^{-1})P \left[ \sum_{k=0}^{2^n - 1}\psi(x_k,t)\ket{k}\right].
\end{equation}
The explicit constructions of quantum circuit for $QFT$, $D$, $QFT^{-1}$ and $P$ are detailed in section \ref{Sec-III}.

\subsection{Analytical Solutions}
Here, we show the analytical solutions for the wave functions in different regions for the three cases namely free space, step potential, double-well potential \cite{peacock2006exact}. In this section the parameter $V_0$ is same as $v$ corresponding to the magnitude of the potential barrier.
\begin{itemize}
\item{\textbf{Free particle :}}
Here, we present analytical solution of free particle to match with our binary representation of wave function. The wave function of particle is
\begin{equation}
 \ket{\psi(x)} = A e^{i\alpha x}+ B e^{-i\alpha x}.    
\end{equation}

\item{\textbf{Step potential }:}
\ Here, the particle is situated in the a step well potential of length ranging from $0$ to $L$ Figure \ref{Fig2.pdf}a  to match with our binary representation of wave function. The wave function of particle is situated in two regions I and II. The region I has potential $V = 0$ and region II has potential $V_0 = 50$. The boundary conditions are 
\begin{equation}
\begin{split}
    V(x) = & \ 0,\ \text{for}\ 0<x< \Delta l_1 \\
           & \ V_0,\ \text{for}\ \Delta l_1<x<L.
\end{split}
\end{equation}
The wave function of particle in region I is
\begin{equation}
 \ket{\psi_{I} (x)} = A sin(\alpha x)\;,
\end{equation}
where $\alpha = \sqrt{\frac{2m}{\hbar^2}E}$, since $V=0$. The wave function of particle in region II is
\begin{equation}
 \ket{\psi_{II} (x)} = B e^{\beta x} + C e^{-\beta x}\;,    
\end{equation}
where where $\beta = \sqrt{\frac{2m}{\hbar^2}[V_0-E]}$.

\item{\textbf{Double square-well potential }:} 
 Here, the particle is situated in the an double square well potential potential of length ranging from $0$ to $L$ Figure \ref{Fig2.pdf}b to match with our binary representation of wave function. The wave function of particle is situated in 4 regions I, II, III and IV. The regions I and III has potential $V_0=50$ and regions II and IV has potential $V=0$. The boundary conditions are 
\begin{equation}
\begin{split}
    V(x) = &\  V_0,\ \text{for} \ 0<x< \Delta l_1\;,\ \Delta l_2<x<\Delta l_3 \\
    & \ 0,\ \text{for}\ \Delta l_1<x<\Delta l_2\;,\ \Delta l_3<x<L.
\end{split}
\end{equation}

\begin{figure}
\centering
\includegraphics[width=\linewidth]{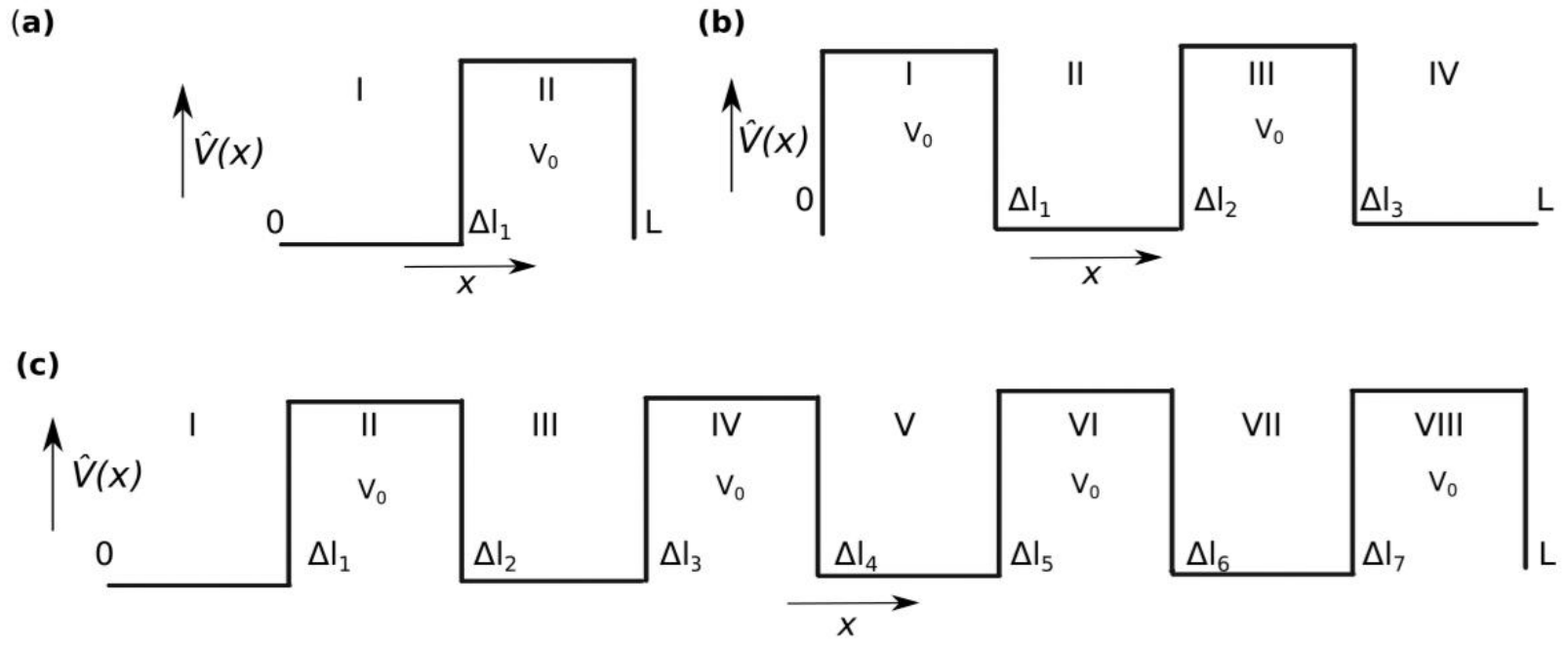}
\caption{Cases (a) and (b)  represent step-well, double square-well schemes, where the height of the barrier $V_0 = 50$. Case (c) represents multi-well potential where $V_0=10$.}
\label{Fig2.pdf}
\end{figure}

The wave functions of particle in regions I and III are 
\begin{equation}
 \ket{\psi_{I} (x)} = A e^{\beta x} + B e^{-\beta x}\;, 
\end{equation}
\begin{equation}
 \ket{\psi_{III} (x)} = D e^{\beta x} + E e^{-\beta x}\;,
\end{equation}
where $\beta = \sqrt{\frac{2m}{\hbar^2}[V_0-E]}$. The wave functions of particle in regions II and IV are
\begin{equation}
 \ket{\psi_{II} (x)} = C sin(\alpha x)\;, 
\end{equation}
\begin{equation}
 \ket{\psi_{IV} (x)} = F sin(\alpha x)\;,
\end{equation}
where $\alpha = \sqrt{\frac{2m}{\hbar^2}E}$, since $V=0$.
\end{itemize}
A similar calculation can be carried out for a particle in multi-well potential shown in Figure \ref{Fig2.pdf}c.

\section{Experimental Procedures \label{Sec-III}}
We investigate quantum simulation of the tunneling process using the IBM’s 5-qubit quantum processor “ibmqx4”, 14-qubit quantum processor “IBM Q 14 Melbourne”, and “ibmqx2". The connectivities between the qubits in both the “ibmqx4" and “IBM Q 14 Melbourne" processors are shown in the supplementary material.

\subsection{Circuit construction for two-qubit simulation}
\begin{figure}
\centering
\includegraphics[width=0.9\linewidth]{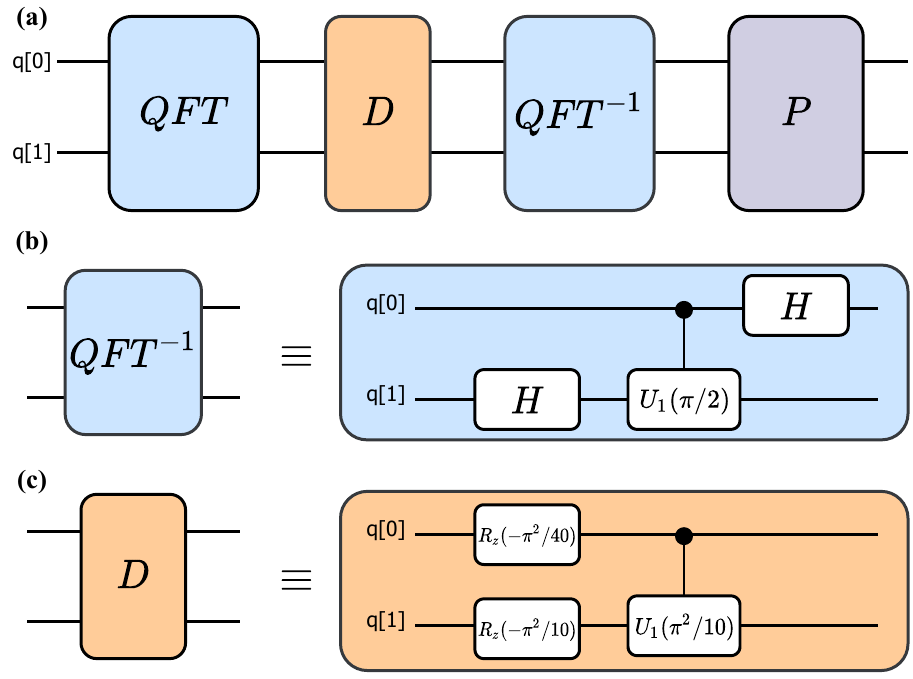}
\caption{ \textbf{Quantum circuit for two qubits simulation}. Case (a): The quantum circuit for one time step (\(\Delta t = 0.1\)) evolution of tunneling Hamiltonian using two qubits is depicted. Case (b): The inverse quantum Fourier transform \(QFT^{-1}\) is built using two Hadamard gates and one controlled-rotation gate \(U_1(\frac{\pi}{2})\). Case (c): The decomposition of diagonal operator $D$ is represented, which requires two single qubit \(R_z\) rotation gates $R_z\left(-\frac{\pi^2}{40}\right)$ \& $R_z\left(-\frac{\pi^2}{10}\right)$ and one controlled-rotation gate \(U_1\left(\frac{\pi^2}{10}\right)\).}
\label{Fig3.pdf}
\end{figure}

The time evolution circuit for one trotter step is shown in Figure \ref{Fig3.pdf}a. Also, the explicit constructions of $QFT^{-1}$, and $D$ are illustrated in Figure \ref{Fig3.pdf}b and \ref{Fig3.pdf}c. The quantum circuit for $QFT^{-1}$ is prepared by two hadamard gates and one controlled-phase gate, 
\begin{equation*}
    QFT^{-1} = H_1 CU_1(\pi/2)_{01} H_0\;,    
\end{equation*} 
where $H_i$ denotes the application of hadamard gate on the $q[i]$ qubit, \(CU_1(\theta)_{ij}\) denotes controlled phase gate of angle $\theta$, where $q[i]$ is the control qubit and $q[j]$ is the target qubit, $U_1(\theta)=[[1,0],[0,e^{i\theta}]]$. For two-qubit simulation, the diagonal elements of $\hat{P}$ are calculated to be \(0,\frac{\pi}{2},\pi,\frac{-\pi}{2}\), see expression \ref{phat}. The quantum circuit for $QFT$ is just the inverse of quantum circuit of \(QFT^{-1}\). Then, the kinetic evolution operator is given by,
\begin{equation}
    \begin{split}
        e^{-i \hat{K} \Delta t} &= (QFT)e^{-i \hat{P}^2 \Delta t}(QFT^{-1}) \\
                                &= (QFT)\Phi_{01} Z_1 Z_0(QFT^{-1}).
    \end{split}
\end{equation}
The diagonal operator $D$ can be expressed as a product of the following operators,
\begin{equation*}
    D= \Phi_{01} Z_1 Z_0\;,
\end{equation*} 
\begin{equation*}
 Z_1 = e^{-i\gamma c_0 \sigma_z^0 \Delta t}, \
 Z_0 = e^{-i\gamma c_1 \sigma_z^1 \Delta t}, \ 
 \Phi_{01}=  e^{-i \gamma c_2 diag(1,1,1,-1)_{01}\Delta t}.
\end{equation*} 
Here, \(\Phi_{01}\) is implemented by \(CU_1 \left(\frac{\pi^2}{10}\right)_{01}\) gate. Here, \(Z_i\) means application of Z operation on $q[i]$
qubit. The constant values are obtained to be \(\gamma = \frac{\pi^2}{8}, c_0 = -1, c_1 = -4, c_2=4\). The time step is taken to be \(\Delta t\) = 0.1 . The quantum circuit for $D$ operator can be given as 

\begin{equation*}
D=R_z\left(\frac{-\pi^2}{40}\right)_0 R_z \left(\frac{-\pi^2}{10}\right)_1 CU_1 \left(\frac{\pi^2}{10} \right)_{01}.
\end{equation*}
Here, \(R_z(\theta)_i\) is the rotation operator about Z-axis by an angle \(\theta\) on the \(q[i]^{th}\) qubit. The \(R_z(\theta)\) can be expressed using standard 'ibmqx4' gates as 

\begin{equation*}
    R_z(\theta) = HU_3\left(\theta = 5,\phi = \frac{-\pi}{2},\lambda=\frac{\pi}{2}\right)H,
\end{equation*} 
Here $U_3$ is defined as follows,
\begin{equation*}
    U_3(\theta,\phi,\lambda)=[[\cos\frac{\theta}{2},-e^{i\lambda}\sin\frac{\theta}{2}],[e^{i\phi}\sin\frac{\theta}{2},e^{i(\lambda + \phi)}\cos\frac{\theta}{2}]]. 
\end{equation*}
The potential operator $\hat{P}$ is prepared by
the application of \(H U_3(\theta=5, \phi = \frac{-\pi}{2}, \lambda =\frac{\pi}{2}) H\) operator. It has been applied on the qubits $q[1]$
and $q[0]$ in case of step potential and double well potential respectively. The controlled phase gates used in the experiment, are directly applied using the codes available in Qiskit.

\subsection{Circuit construction for three-qubit simulation}
\begin{figure}
\centering
\includegraphics[width=0.9\linewidth]{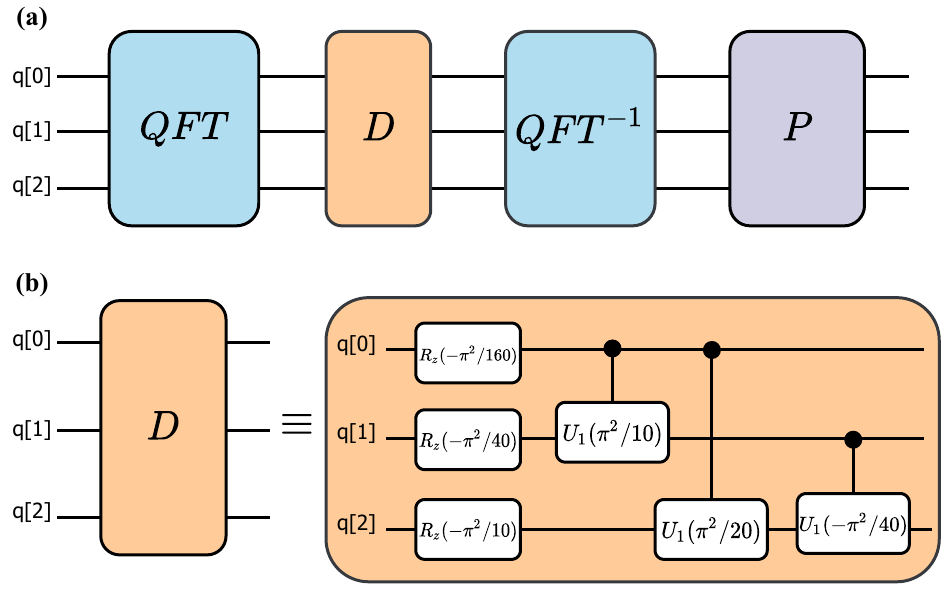}
\caption{ \textbf{Quantum circuit for three-qubit simulation}. Case(a): The quantum circuit for one time
step (\(\Delta t = 0.1\)) evolution of tunneling Hamiltonian using three qubits is depicted. Case(b): The kinetic energy operator $D$ is decomposed into
rotation and controlled-phase gates of \(\frac{-\pi^2}{160}, \frac{-\pi^2}{40}, \frac{-\pi^2}{10}, \frac{\pi^2}{10}, \frac{\pi^2}{20}\).}
\label{Fig10.pdf}
\end{figure}

The explicit constructions of $QFT$, $QFT^{-1}$, $D$ and $P$ are illustrated in Figures \ref{Fig10.pdf}. \(QFT^{-1}\) is prepared by three Hadamard gates and three
controlled-phase gates, 

\begin{equation*}
    (QFT)^{-1} = H_2 CU_1\left(\frac{\pi}{2}\right)_{12} CU_1\left(\frac{\pi}{4}\right)_{02} H_1 CU_1\left(\frac{\pi}{2}\right)_{01} H_0. 
\end{equation*}
The circuit for $QFT$ is just the inverse of \(QFT^{-1}\). For
three-qubit simulation, the diagonal elements of \(P\) are calculated to be \( 0, \frac{\pi}{4}, \frac{\pi}{2}, \frac{3\pi}{4}, \pi,\frac{-\pi}{4}, \frac{-\pi}{2}, \frac{-3\pi}{4}\). 
The decomposition of $D$ operator is as follows,
\begin{equation*}
\begin{split}
   D= & R_z\left(\frac{-\pi^2}{160}\right)_0 R_z\left(\frac{-\pi^2}{40}\right)_1 R_z\left(\frac{-\pi^2}{10}\right)_2 \\ & CU_1\left(\frac{\pi^2}{10}\right)_{01} CU_1\left(\frac{\pi^2}{20}\right)_{02} CU_1\left(\frac{-\pi^2}{40}\right)_{12}. 
\end{split}
\end{equation*}
The diagonal operator $D$ can be expressed as a product of the following operators, 
\begin{equation*}
Z_0 =  e^{-i\gamma c_0 \sigma_z^0 \Delta t},\ 
Z_1 =  e^{-i\gamma c_1 \sigma_z^1 \Delta t},\ 
Z_2 =  e^{-i\gamma c_2 \sigma_z^2 \Delta t},
\end{equation*}
\begin{equation*}
\phi_{01}=  e^{-i\gamma c_3 diag(1,1,1,-1)_{01}\Delta t},\
\phi_{02}=  e^{-i\gamma c_4 diag(1,1,1,-1)_{02}\Delta t},
\end{equation*}
\begin{equation*}
    \phi_{12}= e^{-i\gamma c_5 diag(1,1,1,-1)_{12}\Delta t}.
\end{equation*}
The constant values are obtained to be \( \gamma =\frac{-\pi^2}{32\sqrt{2}}, 
\\c_0 = -1.42,c_1 = -5.66, c_2 = -22.63, c_4 = 11.31, c_5 = -5.66\).  It is to be noted that the time step is taken to be \(\Delta t = 0.1\). The quantum circuit for $D$ operator can be given as 
\begin{equation*}
D = R_z \left(\frac{-\pi^2}{40}\right)_1 R_z \left(\frac{-\pi^2}{10}\right)_2 CU_1 \left(\frac{\pi^2}{10}\right)_{12},    
\end{equation*}
where \(R_z\) is the rotation operator about Z-axis (see Figure \ref{Fig10.pdf}). The potential operator $\hat{P}$ prepared by \( H U_3(\theta,\phi = \frac{-\pi}{2}, \lambda = \frac{\pi}{2}) H \) operator is applied on the qubit $q[0]$ in case of multi well potential. Similarly, the controlled phase gates used here are directly applied using the codes available in Qiskit.

\section{Experimental Results \label{Sec-IV}}
The step potential \cite{young1979studies} is implemented by the quantum operation  $P=e^{-iv\sigma_z^1}\Delta t$ acting on the highest order qubit. Similarly, the double-well potential is implemented by the quantum operation \(P =e^{-iv\sigma_z^0\Delta t}\) acting on the lowest order qubit. In our experiment, we set the parameter $v =0$ for free particle, and $v = 50$ for step potential \cite{young1979studies}, double well potential \cite{cartarius2012model} and $v = 10$ for multi-well potential. The choice of parameter $v$ depends on user performing simulation. One can see for large values of $v$, tunneling will not be possible. As the potential increases, the tunneling probability decreases. We set the time interval $\Delta t = 0.1$, and the mass is the particle is considered as $0.5$.

\begin{figure}
\centering
\includegraphics[width=\linewidth]{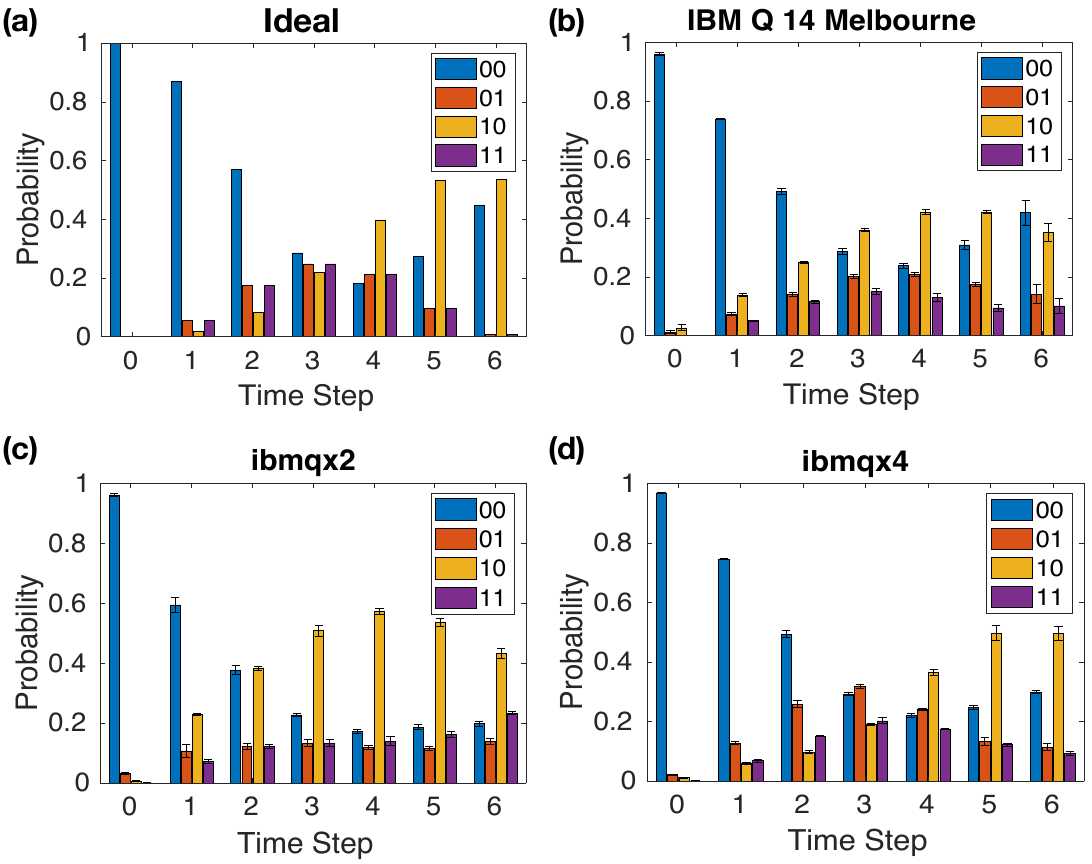}
\caption{\textbf{Free particle($v=0$) probability distribution.} Case (a): Theoretically calculated results for free particle evolution using two-qubit system. Cases (b), (c), (d): Experimentally obtained results for probability distribution of free particle from “IBM Q 14 Melbourne", “ibmqx2" and “ibmqx4" respectively.}
\label{Fig4.pdf}
\end{figure}

\begin{figure*}
\centering
\includegraphics[width=\linewidth]{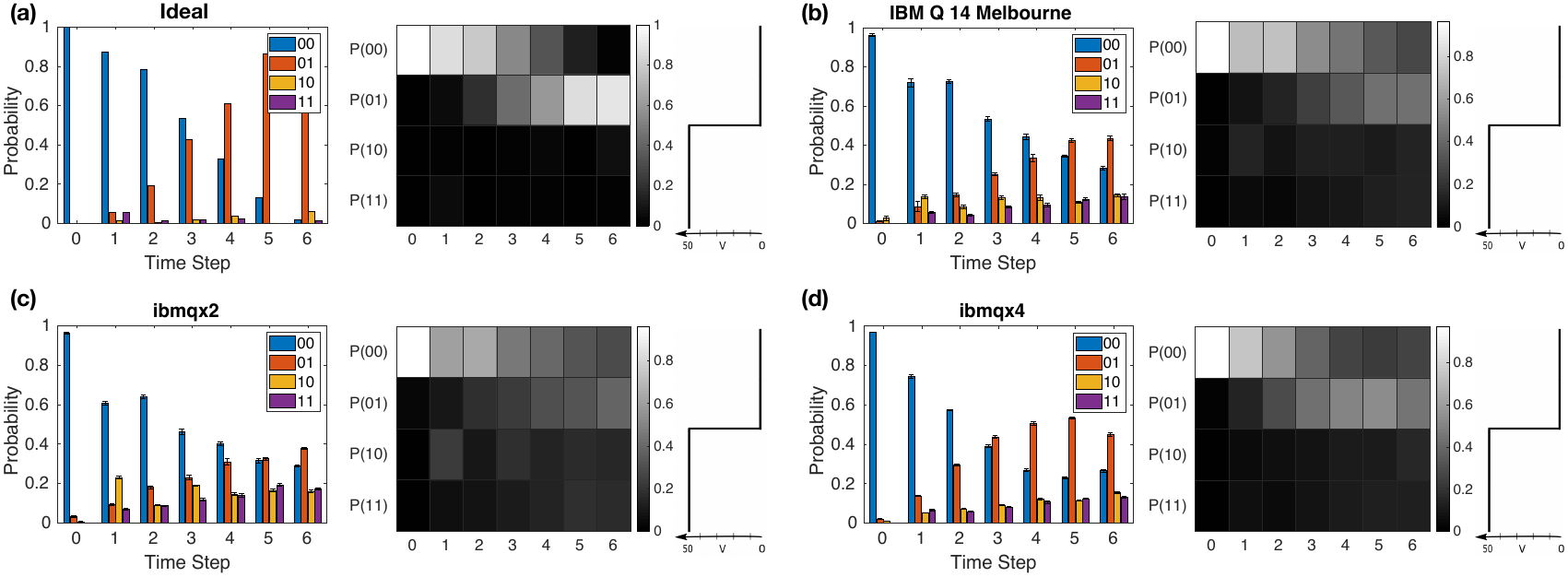}
\caption{\textbf{Particle probability distribution in a step-well potential ($v=50$).} Case (a): Theoretically calculated results and heat map for free particle evolution using two-qubit system. Cases (b), (c), (d): Experimentally obtained results and heat maps for probability distribution of the particle in step well potential using “IBM Q 14 Melbourne", “ibmqx2" and “ibmqx4" respectively.}
\label{Fig5.pdf}
\end{figure*}

In Figure \ref{Fig4.pdf}, the time evolution of a free particle with potential $\hat{V}(x) = 0$ is depicted. Initially, the
particle was confined at $\ket{00}$ state. It can be seen that after a number of time steps, the particle
probability distribution spreads over other basis states.  Both the theoretical and experimental results for free particle evolution are illustrated in Figure \ref{Fig4.pdf}. And both the theoretical and experimental results for particle evolution in step well potential \cite{young1979studies} are depicted in Figure \ref{Fig5.pdf}.
To compare the results obtained from the quantum processor “ibmqx4”, “ibmqx2", “IBM Q 14 Melbourne" and ideal simulators, quantum state tomography is performed. In both the cases, the number of shots taken were $8192$. The
density matrix elements \cite{thekkadath2016direct} for the initial state $\ket{00}$ and the final state after six time steps are depicted
in Figure \ref{Fig7.pdf}. The experimental fidelity \cite{zanardi2007mixed, mendoncca2008alternative, badziag2000local, andersen2013high, ma2009fidelity, oh2002fidelity} for the initial state is $96.79\%$ 
and for the final state after six time step is $93.83\%$.

\begin{figure*}
\centering
\includegraphics[width=\linewidth]{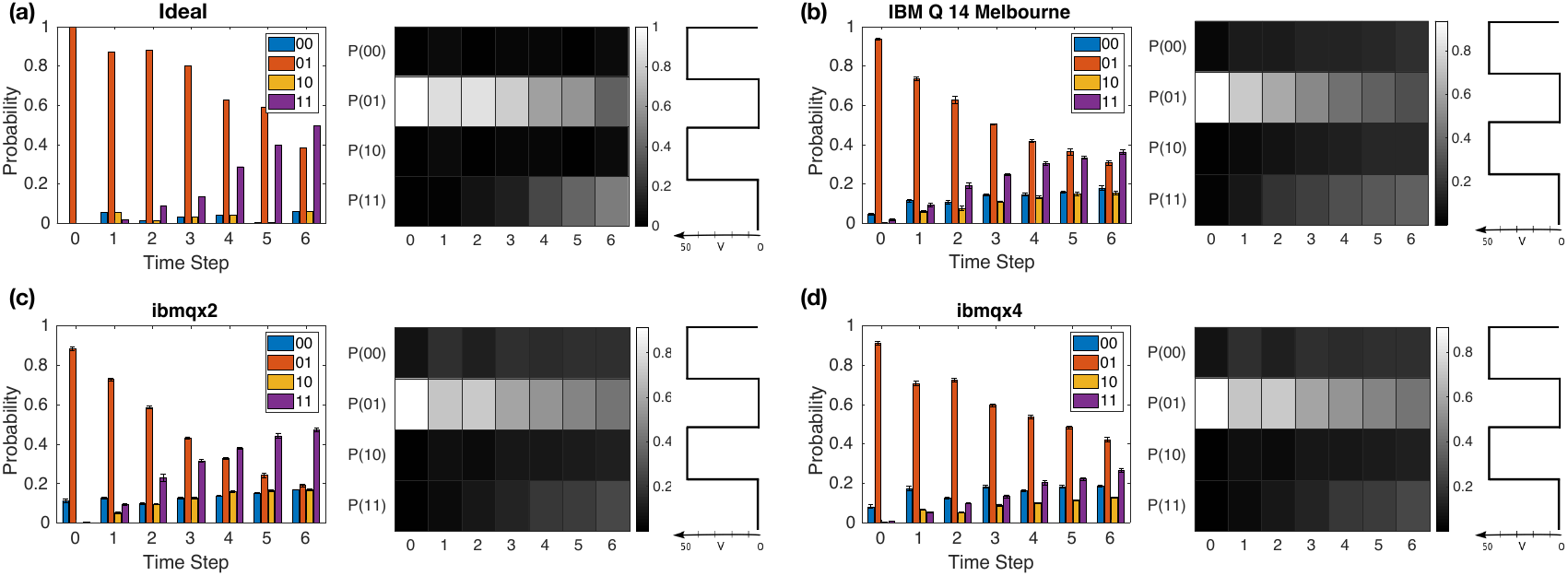}
\caption{\textbf{Particle probability distribution in a double-well potential ($v=50$).} Case (a): Theoretically calculated results and heat map for free particle evolution using two-qubit system. Cases (b), (c), (d): Experimentally obtained results and heat maps for probability distribution of free particle from “IBM Q 14 Melbourne", “ibmqx2" and “ibmqx4" respectively.}
\label{Fig6.pdf}
\end{figure*}

\begin{figure*}
\centering
\includegraphics[scale=0.55]{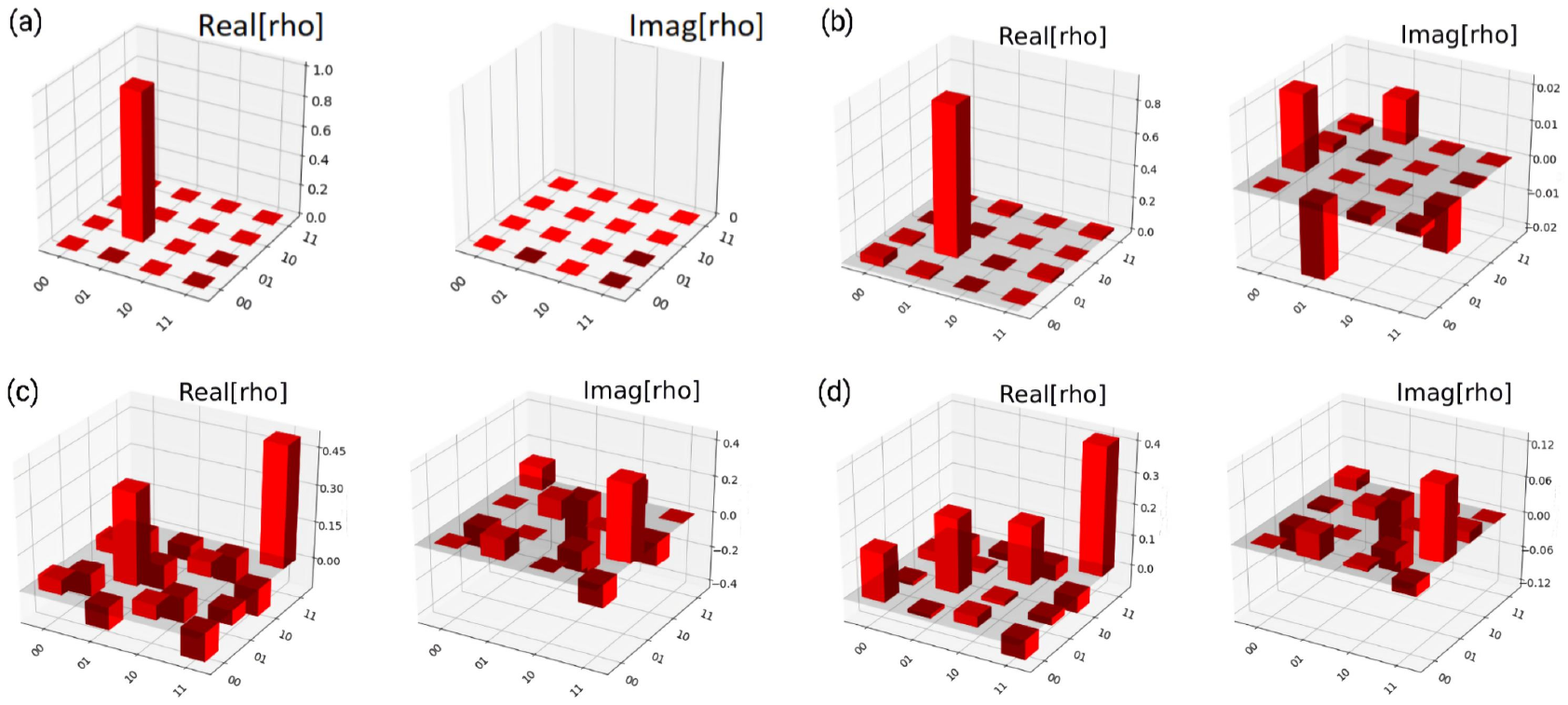}
\caption{Cases (a) and (b) represent theoretical and experimental density matrix elements for the initial
state of the particle in a double well potential. Cases (c) and (d) represent theoretical and experimental
density matrix elements after six time steps.}
\label{Fig7.pdf}
\end{figure*}

In Figure \ref{Fig6.pdf}, we illustrate the tunneling phenomena of a single particle \cite{feng2013experimental} in a finite length step well
potential of $v=50$. The potential barriers are situated at \(\ket{10}\) and \(\ket{11}\) states. At time, $t=0$, the
particle is located at \(\ket{00}\) state, as time evolves we can observe the probability distribution of
the particle spreads to the state \(\ket{01}\). Notably, it can be observed that after six time steps, the
particle has some non-zero probability to be found at classically forbidden region \(\ket{10}\)and \(\ket{11}\).
Which clearly confirms the tunneling of particle through the potential barrier \cite{eckart1930penetration, li2011new}. We perform this.

The tunneling of particle in a double well potential (v=50) \cite{cartarius2012model} is shown in Figure \ref{Fig7.pdf}. \cite{gerritsma2010quantum} The wells
are located at \(\ket{01}\) and \(\ket{11}\) states and the barriers are at \(\ket{00}\) and \(\ket{10}\) states. At time $t=0$, the
particle is confined in the potential well situated at \(\ket{01}\) state, as the time evolves we can clearly
observe the tunneling of particle from the potential well located at \(\ket{01}\) state to the well at \(\ket{11}\)
state. \cite{gerjuoy2005shor}
The theoretical and experimental density matrix elements for the
initial state of the particle and after six time-steps in a double well potential are depicted in Figure \ref{Fig6.pdf}. Both from the theoretical and experimental data shown in Figure \ref{Fig7.pdf}, it is observed that the particle oscillates between the two wells which signifies tunneling through the potential barriers \cite{nag1991boundary, dittrich1985tunneling} at the other two places. To compare the results obtained from the quantum processor “IBM Q 14 Melbourne”, “ibmqx4", “ibmqx2" and the ideal QASM simulator, we perform quantum state tomography \cite{cramer2010efficient, gross2010quantum, christandl2012reliable}.

To perform the experiment, we choose the number of shots to be 8192. The average fidelity \cite{zyczkowski2005average, jozsa1994fidelity} for the initial state $\ket{01}$ is 93.72\%, and for the final state after six time step is 51.8\%. Here the fidelity is calculated by
\begin{eqnarray}
F(\rho, \sigma) = \left[ Tr \sqrt{\sqrt{\rho}\sigma\sqrt{\rho}} \right]^2.
\end{eqnarray}

\begin{figure}
\centering
\includegraphics[width=\linewidth]{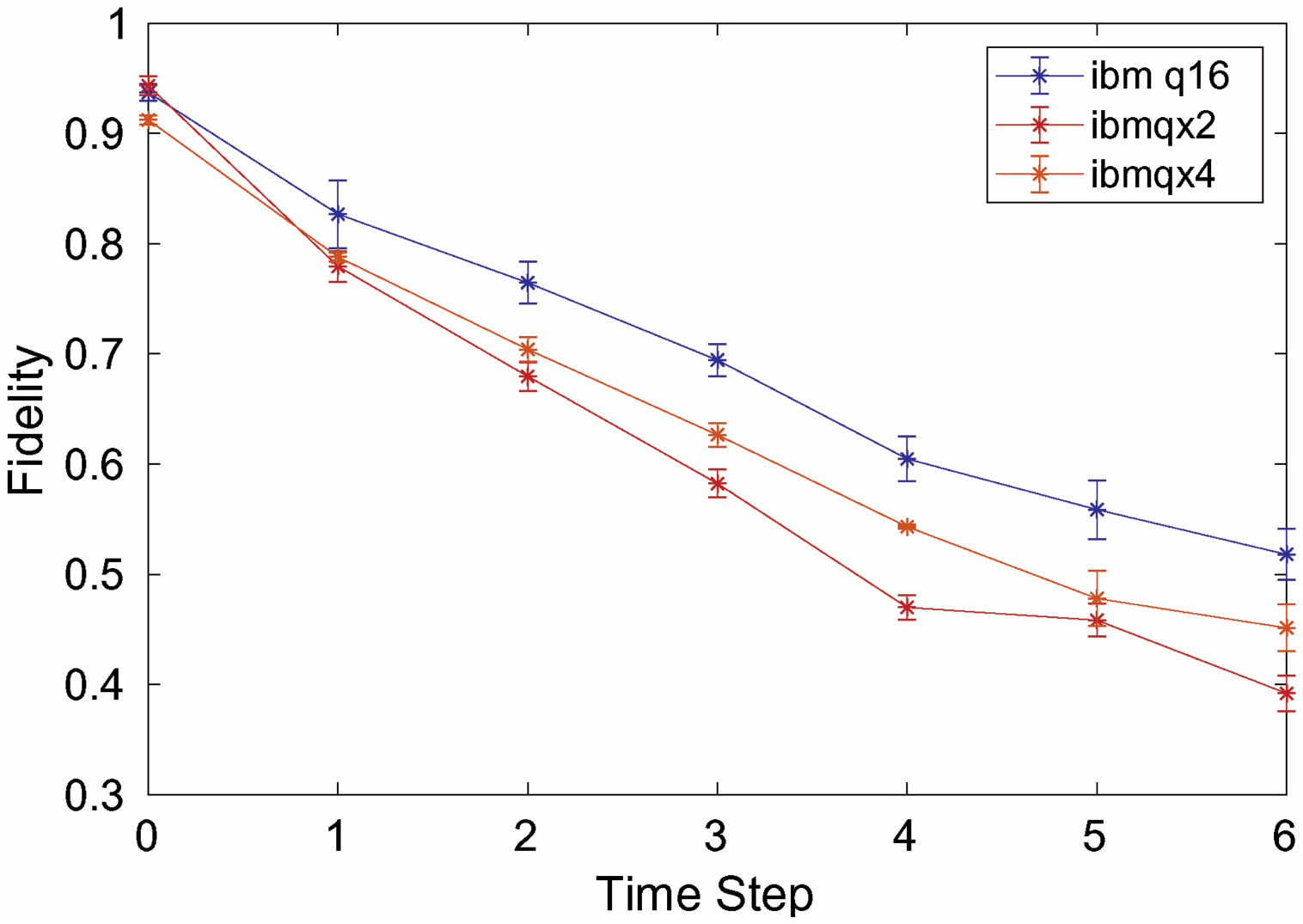}
\caption{The graph shows the variation of fidelity for the evolution of a particle in double-well potential as a function of time steps for different devices.}
\label{Fig9.pdf}
\end{figure}
Here, $\sigma$ represents density matrix obtained by ideal simulation, that
is using Qiskit, and $\rho$ represents density matrix obtained by digital
simulation that is using IBM Q 14 Melbourne, ibmqx4 and ibmqx2 respectively. The variation of fidelity as a function of time steps for different
devices are shown in Figure \ref{Fig9.pdf}.

\begin{figure}
\centering
\includegraphics[width=\linewidth]{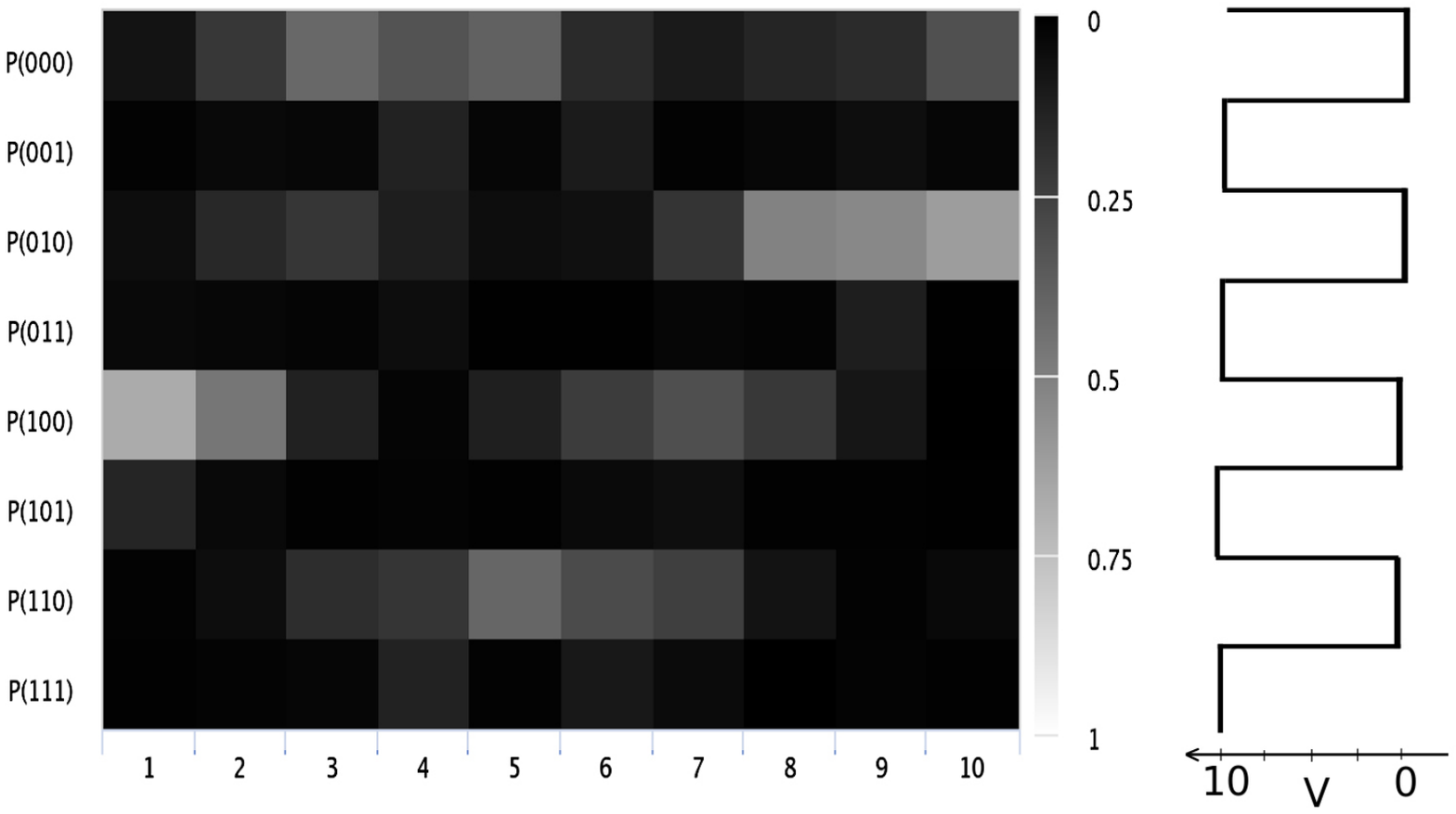}
\caption{ \textbf{Particle probability distributions for ten time steps in a multi-well potential ($v =10$)}. The potential wells are at four sites, \(\ket{000}\), \(\ket{010}\), \(\ket{100}\) and \(\ket{110}\). Initially, the particle is
confined at \(\ket{100}\) state, as the time evolves the particle tunnels through the barriers situated at
$\ket{001}$, $\ket{011}$, $\ket{101}$ and $\ket{111}$ states. After 10 time steps, the particle is most probable to be found
at \(\ket{010}\) state.}
\label{Fig12.pdf}
\end{figure}

Figure \ref{Fig12.pdf} shows the tunneling of the particle in a multi well potential implemented on
the “IBM Q QASM simulator”. The wells are situated at \(\ket{000}\), \(\ket{010}\), \(\ket{100}\) and \(\ket{110}\) positions.
Initially, at time $t=0$, the particle is trapped inside one of the above wells situated at \(\ket{100}\) state, as
time evolves the particle is tunneled through the barriers situated at \(\ket{001}\), \(\ket{011}\),\(\ket{101}\) and \(\ket{111}\)
states. At different time steps, the tunneling of the particle through all the potential barriers is observed. After 10 time steps, the particle is most likely to be found at \(\ket{010}\) state.

\section{Discussions and conclusion \label{Sec-V}}
One of the major applications of a quantum computer in the near future is to simulate complex quantum systems. Here we have experimentally demonstrated the quantum tunneling phenomena of a single particle in a step potential, double-well potential, and multi-well potential. We have designed the equivalent quantum circuit for the time evolution of the given system in the real quantum processors “ibmqx4”, “ibmqx2”, and “IBM Q 14 Melbourne” \cite{buchler2005atomic, byrnes2006simulating}. We have illustrated the tunneling process by running the quantum circuit for six time-steps in two-qubit cases and ten time-steps in three-qubit cases. Here we used one single-qubit gate operation per step for implementing the potential barriers, which drastically reduces the circuit depth while simulating the system over a long time. After comparing the ideal and experimental results, it is concluded that the tunneling process has been properly carried out with IBM’s quantum processors. The average fidelities for the two-qubit system are found to be reduced from nearly 94\% (at $t=0$) to 52\% after six-time steps.The reason for the reduced fidelity is due to the application of a large number of two-qubit gates and the limited coherence time. Future work could include a detailed study of the tunneling process by considering more qubits for higher spatial resolution and simulation of the dynamics by considering more time steps for higher temporal resolution. The basic circuit mechanism used here for demonstrating quantum tunneling can be extended to observe quantum tunneling of multi-particle systems in different shaped potential wells. In the near future, it might be possible to realize the dynamics of complex quantum systems on a quantum computer with high coherence time, high-fidelity gate operation, and lower readout error. We note that the IBM quantum computers that we used are capable of observing the fundamental and interesting quantum phenomenon of tunneling in a small qubit system. Though the particle evolves in continuous space, we have approximated this by considering the particle evolution in discrete space and demonstrated the phenomenon of quantum tunneling on noisy quantum computers. The obtained results implicate the power of a quantum computer and its future promising applications in the field of quantum simulations. The python codes for all the simulations are available on GitHub \cite{code}.

In conclusion, we demonstrate the quantum tunneling of a single particle in various types of potentials like step-well, double-well, and multi-well potentials by means of digital quantum simulation. We used two and three-qubit systems to perform and showcase our work. The quantum tunneling phenomenon can be demonstrated for multi-particle systems as well as for different types of potentials by using this simulation technique. One can probe the quantum to classical transition \cite{jeong2014coarsening, li2012experimental} in a double-well potential in this approach and extend for more qubit systems as well.

\section{Acknowledgements}
N.N.H., B.D., N.L.K. are grateful to IISER Kolkata for cordial hospitality. B.K.B. is financially supported by Institute Fellowship. The authors are extremely grateful to IBM team and IBM QE project. The discussions and opinions developed in this paper are only those of the authors and do not reflect the opinions of IBM or IBM QE team. 

\bibliography{bibliography.bib}

\pagebreak
\clearpage

\end{document}